\input{aipcheck}

\documentclass[final]{aipproc}

\usepackage{amssymb,amsmath}
\usepackage{multirow}
\usepackage{verbatim}

\layoutstyle{8x11single}

\def\beq{\begin{equation}}
\def\eeq{\end{equation}}
\def\beqa{\begin{eqnarray}} 
\def\eeqa{\end{eqnarray}}

\def\mn{{\mu\nu}}

\begin{document}

\title{Modified theories of gravity with non-minimal curvature-matter coupling}

\classification{04.50.Kd, 98.80.-k, 98.80.Jk}
\keywords      {Modified theories of gravitation, Energy Conditions, Dolgov-Kawasaki criterion.}

\author{Orfeu Bertolami\footnote{Also at Instituto de Plasmas e Fus\~{a}o Nuclear, Instituto Superior T\'ecnico,
Lisboa. E-mail address: orfeu@cosmos.ist.utl.pt}}{
  address={Instituto Superior T\'ecnico, Departamento de F\'\i sica, \\
Av. Rovisco Pais, 1049-001 Lisboa, Portugal}
}

\author{Miguel Carvalho Sequeira}{
  address={Instituto Superior T\'ecnico, Departamento de F\'\i sica, \\
Av. Rovisco Pais, 1049-001 Lisboa, Portugal}
}

\begin{abstract}
In this contribution one examines the generalization of the $f(R)$ theories of gravity where one introduces a non-minimal coupling between curvature and matter. This model has new and interesting features. 
However, as any modified gravity theories, these may give origin to anomalies which might turn the theory physically meaningless. In this respect, one undertakes a study on the energy conditions and the Dolgov-Kawasaki criterion specific of this new model.

\end{abstract}

\maketitle

\section{Introduction}

The so-called $f(R)$ theories of gravity  have been recently receiving great attention due to the possibility they open to account for the late time accelerated expansion of the universe without the need of explicit additional degrees of freedom  and also for replacing dark matter at galactic level and beyond (see, for instance, Refs. \cite{Nojiri06,Sotiriou:2008rp} for reviews).

In this contribution one studies the $f(R)$ theories with non-minimal coupling to matter from the point of view of energy condition and stability. The results presented here mirror the ones obtained in Ref. \cite{Bertolami:2009cd}.

More concretely, one is interested in a model with a non-minimal coupling between the Ricci scalar, $R$, and the matter Lagrangian density, ${\cal L}_{m}$ \cite{Bertolami:2007} (see also Ref. \cite{Bertolami:2008zh} for a recent review). The action of interest has the following form,
\begin{equation} \label{action}
S=\int \left[ \frac{k}{2}f_1(R)+f_2(R){\cal L}_m \right] \sqrt{-g}d^4x\,,
\end{equation}
where $k$ is a coupling constant, $f_i(R)$ (with $i=1,2$) are arbitrary functions of $R$ and ${\cal L}_{m}$. One sets $k=1$ hereafter. The second function, $f_2(R)$, is usually considered to have the following form,

\begin{equation} \label{f2}
f_2(R)=1+\lambda\varphi_2(R)
\end{equation}
where $\lambda$ is a constant and $\varphi_2$ is another function of $R$.

Varying action (\ref{action}) with respect to the metric $g_{\mu \nu }$
yields the field equations
\begin{equation}\label{fieldEq}
(f'_1+2 {\cal L}_m f'_2)R_{\mu \nu }-\frac{1}{2}f_1g_{\mu \nu }-\Delta_{\mu\nu}(f'_1+2{\cal L}_m f'_2)= f_2T_{\mu\nu }
 \,,
\end{equation}
where the prime denotes the differentiation with respect to $R$, $\Delta_{\mu\nu}=\nabla_\mu \nabla_\nu - g_{\mu\nu}\Box$ and $\Box = g^{\mu\nu}\nabla_\mu \nabla_\nu$. The matter energy-momentum tensor is defined as usual
\begin{equation}
T_{\mu \nu}=-\frac{2}{\sqrt{-g}}{\delta(\sqrt{-g}{\cal L}_m)\over \delta(g^{\mu\nu})} \,. \label{defSET}
\end{equation}

This theory has some interesting features. The most intriguing is that the non-minimal curvature-matter coupling implies a violation of the conservation equation of the energy-momentum tensor which in turn implies that the motion of a point like particle is non-geodesic \cite{Bertolami:2007}. Actually, this issue has been a lively issue of discussion in the literature \cite{Sotiriou08,Sotiriou:2008it,Bertolami:2008ab,Puetzfeld:2008xu,Bertolami:2008im}. These modified gravity models have also been examined from the point of view of their impact on stellar stability \cite{Bertolami:2007vu} and its ability to account for the rotation curves of galaxies \cite{Bertolami:2009ic}.

In this contribution one analyzes these modified models of gravity with non-minimal coupling between curvature and matter from the point of view of the energy conditions and their stability under the Dolgov-Kawasaki criterion. As will be discussed, the energy conditions are necessary to ensure that energy-momentum tensor is physically consistent, whereas the Dolgov-Kawasaki criterion is crucial to ensure the stability of the theory \cite{Faraoni}.




\section{Energy Conditions}\label{energyConditions}

\subsection{The Raychaudhuri Equation}\label{raychaudhuriEq}

The origin of the so-called Strong Energy Conditions (SEC) and of the Null Energy Condition (NEC)  is the Raychaudhuri equation together with the requirement that the gravity is attractive for a space-time manifold endowed with a metric $g_{\mu\nu}$ (see e.g. Ref. \cite{Hawking:1973uf} for a discussion). From the Raychaudhuri equation one has that the temporal variation of the expansion of a congruence defined by the vector field $u^{\mu}$ is given for the case of a congruence of timelike geodesics by
\begin{equation}\label{rayTimelike}
\frac{d \theta}{d\tau}=-\frac{1}{3}\theta^2-\sigma_{\mu\nu}\sigma^{\mu\nu}+\omega_{\mu\nu}\omega^{\mu\nu}-R_{\mu\nu}u^\mu u^\nu ~~,
\end{equation}
where $R_{\mu\nu}$, $\theta$, $\sigma_{\mu\nu}$ and $\omega_{\mu\nu}$ are the Ricci tensor, the expansion parameter, the shear and the rotation tensors associated to the congruence. In the case of a congruence of null geodesics, defined by the vector field $k^\mu$, one has to replace the first and last terms of Eq. (\ref{rayTimelike}) by $-\frac{1}{2}\theta^2$ and $-R_{\mu\nu}k^\mu k^\nu$, respectively.

It is important to realise that the Raychaudhuri equation is derived from purely geometric principles and is independent from the gravity theory under consideration. Thus, imposing the requirement for attractive gravity, one is able to express the SEC and the NEC in terms of geometric quantities. Indeed, since $\sigma_{\mu\nu}\sigma^{\mu\nu}\geq0$, one has, from Eq. (\ref{rayTimelike}), that the conditions for gravity to remain attractive ($d\theta/d\tau<0$) are given by
\begin{eqnarray}
R_{\mu\nu}u^\mu u^\nu\geq 0 \text{\qquad \qquad SEC} \label{SEC} \\ 
R_{\mu\nu}k^\mu k^\nu \geq 0 \text{\qquad \qquad NEC}\label{NEC}
\end{eqnarray}
for any hypersurface of orthogonal congruences. Of course, the requirement for attractive gravity may not hold at all instances. Indeed, a repulsive interaction is what is required to avoid singularities as well as to achieve inflationary conditions, and to account the observed accelerated expansion of the universe.
From the inequalities (\ref{SEC}) and (\ref{NEC}), one realises that the connection with the gravity theory comes from the fact that, in order to obtain the energy conditions in terms of the matter-energy variables, one needs the field equations to relate the Ricci tensor with the energy-momentum tensor. 

For instance, if one uses Einstein's field equations into Eqs. (\ref{SEC}) and (\ref{NEC}) one obtains  the well known SEC and NEC: 
\begin{eqnarray}
\rho+3p\geq 0 \text{\qquad \qquad SEC} \label{SECGR} \\ 
\rho+p \geq 0 \text{\qquad \qquad NEC} \label{NECGR}
\end{eqnarray}
where one considers a energy-momentum tensor for a perfect fluid with energy density $\rho$ and pressure $p$, which is given by
\begin{equation}\label{perfectEMTensor}
T_\mn=(\rho + p)u_\mu u_\nu - p g_\mn \, .
\end{equation}
%
\subsection{Effective Energy-Momentum Tensor}

In what follows, one generalizes these conditions to the gravity model with non-minimal curvature-matter coupling, a procedure developed in Ref. \cite{Santos} for $f(R)$ theories.

Rewriting Eq. (\ref{fieldEq}) in order to make explicit the Einstein tensor, $G_\mn$, one gets
\begin{equation}\label{fieldEq2}
G_\mn=\hat k \left(\hat T_\mn + T_\mn \right)\,,
\end{equation}
where it has been defined an effective energy-momentum tensor 
\begin{equation}\label{effectiveEMTensor}
\hat T_\mn=\frac{1}{2}\left(\frac{f_1}{f_2} - \frac{f'_1+2{\cal L}_m f'_2}{f_2}R \right)g_\mn+\frac{1}{f_2}\Delta_\mn\left(f'_1+2{\cal L}_m f'_2\right)\,,
\end{equation}
and an effective coupling
\begin{equation}
\hat k=\frac{f_2}{f'_1+2{\cal L}_m f'_2} \, .
\end{equation}

From the last definition one realizes that, in order to keep gravity attractive, in addition to conditions Eqs. (\ref{SEC})-(\ref{NEC}), $\hat k$ has to be positive, from which follows the additional condition
\begin{equation}\label{couplingIneq}
\frac{f_2}{f'_1+2{\cal L}_m f'_2} >0 \, .
\end{equation}

Notice that this condition is independent from the ones derived from the Raychaudhuri equation (Eqs. (\ref{SEC}) and (\ref{NEC})). While the latter is derived directly from geometric principles, the former is related with the fairly natural definition of an effective gravitational coupling.

By defining an effective energy density and an effective pressure, one can rewrite the effective energy-momentum tensor (Eq. (\ref{effectiveEMTensor})) in the form of a perfect fluid (Eq. (\ref{perfectEMTensor})). However, in order to deal with the higher order derivatives present in Eq. (\ref{effectiveEMTensor}), one has to specify the metric of the space-time manifold of interest. Since one is concerned with cosmological applications, the Robertson-Walker (RW) metric is a natural choice. In what follows, one considers the homogeneous and isotropic flat RW metric with the signature $(+,-,-,-)$,
\begin{equation}\label{RWmetric}
ds^2=dt^2-a^2(t) ds^2_3 \, ,
\end{equation}
where $ds^2_3$ contains the spacial part of the metric and $a(t)$ is the scale factor.

Using this metric, the effective energy density is given by
\begin{equation}
\hat \rho=\frac{1}{2}\left(\frac{f_1}{f_2}-\frac{f'_1+2{\cal L}_m f'_2}{f_2}R\right)-3H\frac{f''_1+2{\cal L}_mf''_2}{f_2}\dot R\, ,
\end{equation}
while the effective pressure is given by
\begin{eqnarray}
\hat p=-\frac{1}{2}\left(\frac{f_1}{f_2}-\frac{f'_1+2{\cal L}_m f'_2}{f_2}R\right)+ 
 (\ddot R+2H\dot R)\frac{f''_1+2{\cal L}_mf''_2}{f_2}+
\frac{f'''_1+2{\cal L}_m f'''_2}{f_2}\dot R^2 \,,
\end{eqnarray}
where the dot refers to derivative with respect to cosmic time.

\subsection{Strong and Null Energy Conditions}

Given the above results, one is able to obtain explicit expressions for the energy conditions. Using Eq. (\ref{fieldEq2}), the SEC and the NEC given by Eqs. (\ref{SEC}) and (\ref{NEC}) become

\begin{eqnarray}
\hat k \left(\hat T_\mn +T_\mn\right)u^\mu u^\nu -\frac{1}{2}\hat k\left(\hat T+T \right) \geq 0 \text{\qquad \qquad SEC} \label{SEC3} \\
\hat k \left(\hat T_\mn +T_\mn\right)k^\mu k^\nu \geq 0 \text{\qquad \qquad NEC} \label{NEC3}
\end{eqnarray}
where $T$ and $\hat T$ are the traces of $T^\mn$ and $\hat T^\mn$ respectively. 

Using Eq. (\ref{couplingIneq}) and assuming a perfect fluid as the matter source, one obtains for the SEC
\begin{eqnarray}\label{modifiedSEC}
\rho +3p - \left(\frac{f_1}{f_2}-\frac{f'_1+2{\cal L}_m f'_2}{f_2}R \right)
 +3(\ddot R + H\dot R) \frac{f''_1+2{\cal L}_m f''_2}{f_2}+
 3\frac{f'''_1+2{\cal L}_m f'''_2}{f_2}\dot R^2 \geq 0 \,,
\end{eqnarray}
and for the NEC
\begin{equation}\label{modifiedNEC}
\rho+p+(\ddot R -H\dot R)\frac{f''_1+2{\cal L}_m f''_2}{f_2}+\frac{f'''_1+2{\cal L}_m f'''_2}{f_2}\dot R^2 \geq 0 \,.
\end{equation}

Notice that despite the fact that the above results are obtained directly from the Raychaudhuri equation, similar results could have been obtained by applying the transformations $\rho \rightarrow \rho + \hat\rho$ and $p\rightarrow p+ \hat p$ to the SEC and NEC of general relativity.

Naturally, setting $f_2=1$ in Eqs. (\ref{modifiedSEC}) and (\ref{modifiedNEC}) one recovers the results of Ref. \cite{Santos}. Setting $f_1=R$ and $f_2=1$ one recovers SEC and NEC of general relativity (Eqs. (\ref{SECGR}) and (\ref{NECGR})).

It is interesting to note that if one demands that gravity is repulsive, the signs of the inequalities (\ref{SEC3}) and (\ref{NEC3}) would be the same, since the signs of the conditions (\ref{SEC}), (\ref{NEC}) and (\ref{couplingIneq})
would change, and thus the above results for the SEC and the NEC would remain unaltered.

\subsection{Dominant and Weak Energy Conditions}

Extending the results obtained above, one can obtain the Dominant Energy Condition (DEC) and the Weak Energy Condition (WEC) by applying the transformations $\rho \rightarrow \rho + \hat \rho$ and $p\rightarrow p + \hat p$ to the DEC and WEC of general relativity since it is natural to assume that the energy conditions will not be violated when one changes from the Jordan frame to the Einstein frame (see, for instance, Ref. \cite{Atazadeh08}).

Thus, one gets for the DEC
\begin{eqnarray}\label{modifiedDEC}
\rho - p + \left(\frac{f_1}{f_2}-\frac{f'_1+2{\cal L}_m f'_2}{f_2}R\right) -
(\ddot R + 5H \dot R)\frac{f''_1+2{\cal L}_m f''_2}{f_2} -
\frac{f'''_1+2{\cal L}_m f'''_2}{f_2} \dot R^2 \geq 0
\end{eqnarray}
and for the WEC
\begin{equation}\label{modifiedWEC}
\rho + \frac{1}{2}\left(\frac{f_1}{f_2}-\frac{f'_1+2{\cal L}_m f'_2}{f_2}R\right) - 3H\frac{f''_1+2{\cal L}_m f''_2}{f_2}\dot R \geq 0 \,.
\end{equation}

As one may easily check that, for $f_2=1$ one obtains the results of Ref. \cite{Santos}. If one further sets $f_1=R$,  Eqs. (\ref{modifiedDEC}) and (\ref{modifiedWEC}) yield $\rho-p\geq0$ and $\rho\geq0$, as expected. As it is well known from GR these conditions guarantee that the sound velocity does not exceed the speed of light, and that the energy is positive, respectively.

\subsection{Energy Conditions for a Class of Models}

To get some insight on the meaning of the above energy conditions, one applies then to a specific type of models where $f_{1,2}$ are given by
\begin{align}
f_1(R) &=R+\epsilon R^n \nonumber \,, \\
\varphi_2(R) &=R^m \,. \label{firstModels}
\end{align}
\noindent Assuming a flat RW geometry, Eq.  (\ref{RWmetric}), the energy conditions can be written in the following way,
\begin{equation}\label{4EnergyConditions}
\frac{\hat \epsilon |R|^n}{1+\hat\lambda |R|^m}\left( a-\alpha_n-\frac{2\hat\lambda {\cal L}_m}{\hat \epsilon}\alpha_m|R|^{m-n} \right) \geq b
\end{equation}
where $a$, $b$ and $\alpha_{n,m}$ depend on the energy condition under study and one defines $\hat \epsilon=(-1)^n\epsilon$ and $\hat \lambda=(-1)^m\lambda$ due to the fact that for a RW metric one has $R<0$. For the SEC, one finds
\begin{equation}
a^{SEC}=-1\,, \ \ \  b^{SEC}=-(\rho+3p)\, ,
\end{equation}
\begin{eqnarray}
\alpha^{SEC}_n= -n\left[1+3(n-1)(\ddot R+H\dot R)R^{-2}+
 3(n-1)(n-2)R^{-3}\dot R^2 \right].
\end{eqnarray}

For the NEC, one obtains
\begin{equation}
a^{NEC}=0\,, \ \ \  b^{NEC}=-(\rho+p)\, ,
\end{equation}
\begin{equation}
\alpha^{NEC}_n= -n (n-1) \left[(\ddot R-H\dot R)R^{-2}+(n-2)R^{-3}\dot R^2 \right]\, .
\end{equation}

For the DEC, one has
\begin{equation}
a^{DEC}=1\,, \ \ \  b^{DEC}=-(\rho-p)\, ,
\end{equation}
\begin{eqnarray}
\alpha^{DEC}_n=n\left[1+(n-1)(\ddot R+5H\dot R)R^{-2}+
(n-1)(n-2)R^{-3}\dot R^2\right]\, .
\end{eqnarray}

Finally, for the WEC, one gets
\begin{equation}
a^{WEC}=\frac{1}{2}\,, \ \ \  b^{WEC}=-\rho\, ,
\end{equation}
\begin{equation}
\alpha^{WEC}_n=n\left[\frac{1}{2}+3(n-1)HR^{-2}\dot R\right]
\, .
\end{equation}

Given these definitions, the study of all the energy conditions can be performed by satisfying the inequality Eq. (\ref{4EnergyConditions}). In Table \ref{EnergyConditionsTable} it is presented the energy conditions for models of type Eq. (\ref{firstModels}). One realizes that all the energy conditions depend on the couplings of the model and on the space-time under consideration. The balance between the couplings and the space-time parameters and their evolution along the history of the universe will determine whether the energy conditions are satisfied or not. 

Furthermore, writing the derivatives of $R$ and hence the $\alpha_{n}$ parameters in terms of the deceleration (q), jerk (j) and snap (s) parameters, one is able to study the energy conditions from an observational perspective. Indeed, setting the estimated values of $q$, $j$ and $s$ from observation, one can impose bounds on the couplings of the gravity model (see discussion on Ref. \cite{Bertolami:2009cd}).

%
\begin{table}[t]
\begin{tabular}{|c c|c|c|}
\hline
$ $& & $\hat\epsilon >0$ & $\hat\epsilon <0$\\
\hline
\multicolumn{2}{|c|}{$\hat\lambda >0$} & $\frac{2\hat\lambda{\cal L}_m}{\hat\epsilon}\leq\frac{a-\alpha_n}{\alpha_m}|R|^{n-m}$ &$\frac{2\hat\lambda{\cal L}_m}{|\hat\epsilon|}\leq\frac{\alpha_n-a}{\alpha_m}|R|^{n-m}$ \\
\hline
\multirow{2}{*}{$\hat\lambda<0$} &$1-|\hat\lambda||R|^m>0$ &$\frac{2|\hat\lambda|{\cal L}_m}{\hat\epsilon}\geq\frac{\alpha_n-a}{\alpha_m}|R|^{n-m}$  &$\frac{2|\hat\lambda|{\cal L}_m}{|\hat\epsilon|}\geq\frac{a-\alpha_n}{\alpha_m}|R|^{n-m}$ \\
\cline{2-4}
&$1-|\hat\lambda||R|^m<0$ &$\frac{2|\hat\lambda|{\cal L}_m}{\hat\epsilon}\leq\frac{\alpha_n-a}{\alpha_m}|R|^{n-m}$ & $\frac{2|\hat\lambda|{\cal L}_m}{|\hat\epsilon|}\leq\frac{a-\alpha_n}{\alpha_m}|R|^{n-m}$\\
\hline
\end{tabular}
\caption{SEC, NEC, DEC and WEC expressed by Eq. (\ref{4EnergyConditions}) with $b=0$ and where it is assumed that $\alpha_m>0$. For $\alpha_m<0$ one has just  to change the direction of the inequalities.}
\label{EnergyConditionsTable}
\end{table}
%

\section{Generalized Dolgov-Kawasaki Instability}\label{massInstability}

Modified gravity theories may be, in contrast to general relativity, intrinsically unstable. This difference is due to the fact that the modified field equations (Eq. (\ref{fieldEq})) are of fourth order, with their trace being a dynamical equation for $R$, whereas the ones from general relativity are of second order and their trace gives an algebraic equation for $R$. In order for the dynamical field $R$ to be stable its ``mass'' must be positive, a requirement usually referred to as Dolgov-Kawasaki stability criterion.

The Dolgov-Kawasaki criterion was studied in the context of $f(R)$ theories in Ref. \cite{Nojiri03} and its generalization for the models under study  and can be expressed as \cite{Bertolami:2009cd,Faraoni}
\begin{equation}\label{inequalityDK}
f_1''(R)+2\lambda{\cal L}_m\varphi''_2(R)\geq0,
\end{equation}
where $\varphi_2(R)$ is defined by Eq. (\ref{f2}) and $f_1(R)$ is written as $f_1(R)=R+\epsilon\varphi_1(R)$, with $\epsilon$ being a constant. 

For models such as the ones defined by Eq. (\ref{firstModels}), the inequality Eq. (\ref{inequalityDK}) becomes
\begin{equation}\label{inequality1}
\hat\epsilon n(n-1) |R|^{n-2} +2\hat\lambda{\cal L}_m m(m-1)|R|^m\geq0 \,,
\end{equation}
where $\hat\epsilon$ and $\hat\lambda$ are defined by,
\begin{equation}
\hat\epsilon={\left\{\begin{array}{cc} (-1)^n\epsilon,&\mbox{ if } R< 0\\
\epsilon, & \mbox{ if } R>0\end{array}\right.}\,, \ \ \
\hat\lambda={\left\{\begin{array}{cc} (-1)^m\lambda,&\mbox{ if } R< 0\\
\lambda, & \mbox{ if } R>0\end{array}\right.}\,.
\end{equation}

Once again, one notices that this inequality can be written as Eq. (\ref{4EnergyConditions}) with the parameters,
\begin{equation}
a^{DK}=0\,,\ \ \  b^{DK}=0\, ,
\end{equation}
\begin{equation}
\alpha^{DK}_n=-\frac{n(n-1)}{R^2}(1+\hat\lambda|R|^m)\, .
\end{equation}

Thus, the results presented in Table \ref{EnergyConditionsTable}, for the cases with $1+\hat\lambda|R|^m>0$, stand for condition Eq. (\ref{inequality1}) too. However, it is important to realize that although the energy conditions and the Dolgov-Kawasaki criterion yield the same type of inequalities, they are independent from each other. From Table \ref{EnergyConditionsTable}, one can see that, if the factors $\frac{a-\alpha_n}{\alpha_m}$  are the same, the various conditions (the four energy conditions and the Dolgov-Kawasaki criterion) are degenerate.

For most of the cases, the viability of the model with respect to the Dolgov-Kawasaki instability criterion will depend not only on the value of the constants $\epsilon$ and $\lambda$ (which may be further constrained from the solar system observations \cite{Bertolami:2008im}), but also on the space-time metric under consideration. Furthermore, since $\alpha^{DK}_n/\alpha^{DK}_m>0$, one can see that models for which $\hat\lambda>0$ and $\hat\epsilon>0$, the Dolgov-Kawasaki criterion is always satisfied independently of the values of $\epsilon$, $\lambda$ and $R$ (note that as $\alpha^{DK}_m<0$, the direction of the inequalities in Table \ref{EnergyConditionsTable} is reversed).

For models where $m=n$ the inequality Eq. (\ref{inequality1}) implies, $\epsilon + 2\lambda{\cal L}_m \geq 0$.

\section{Conclusions}\label{conclusions}

In this contribution, one has examined the energy conditions and the Dolgov-Kawasaki stability criterion for $f(R)$ theories with a non-minimal curvature-matter coupling. It is shown that the generalized SEC and NEC can be obtained directly from the the SEC and NEC of general relativity through the transformations $\rho\rightarrow\rho+\hat \rho$ and $p\rightarrow p + \hat p$, where the quantities $\hat \rho$ and $\hat p$ contain the physics of the higher-order terms in $R$. The same transformations were applied in order to obtain the generalized DEC and WEC. Conditions to keep the effective gravitational coupling positive and gravity attractive (Eq. (\ref{couplingIneq})) were also obtained.  One stresses that this method, and in particular Eqs. (\ref{couplingIneq}), (\ref{modifiedSEC}), (\ref{modifiedNEC}), (\ref{modifiedDEC}) and (\ref{modifiedWEC}) are quite general and can be used to study the physical implications of any $f(R)$ model with non-minimal curvature-matter coupling

Though the application of the encountered energy conditions, as well as the Dolgov-Kawasaki criterion, to a specific kind of models, one realizes that it is possible to unified then in one single inequality with specific parameters. As expected, all generalized energy conditions and the Dolgov-Kawasaki criterion for a suitable energy-momentum tensor are strongly dependent on the geometry. This fact may impose further constrains on the theory since it restricts the Ricci scalar, which might turn out to be unphysical for some models.



\end{document}